\include{graphicsx} 

\documentclass[12pt,preprint]{aastex}

\usepackage{emulateapj5}

\slugcomment{To appear in the November 20, 2003 issue of the Astrophysical Journal Letters}

\shorttitle{Mid-IR AO Imaging of AC Her}
\shortauthors{Close et al.}

\begin{document}


\title{Mid-Infrared Imaging of the Post-AGB Star AC Herculis with the MMT Adaptive Optics System{\footnotemark}}

\footnotetext{The results presented here made use of the of MMT Observatory, a facility jointly operated by the University of Arizona and the Smithsonian Institution.}


\author{Laird M. Close$^1$, Beth Biller$^1$, William F. Hoffmann$^1$, Phil M. Hinz$^1$, John H. Bieging$^1$, Francois Wildi$^1$, Michael Lloyd-Hart$^1$, Guido Brusa$^{1,2}$, Don Fisher$^1$, Doug Miller$^1$, Roger Angel$^1$}

\email{lclose@as.arizona.edu}

\affil{$^1$Steward Observatory, University of Arizona, Tucson, AZ 85721}
\affil{$^2$Arcetri Observatory, Universita degli Studi di, Firenzi, I-50125 Italy}



\begin{abstract} 

 We utilized the unique 6.5m MMT deformable secondary adaptive optics
system to produce high-resolution (FWHM=0.3$\arcsec$), very high
Strehl mid-infrared (9.8, 11.7 \& 18 $\mu$m) images of the post-AGB
star AC Her. The very high ($98\pm2\%$) Strehls achieved with Mid-IR
AO led naturally to an ultra-stable PSF independent of airmass,
seeing, or location on the sky. We find no
significant difference between AC Her's morphology and our unresolved
PSF calibration stars ($\mu$ UMa \& $\alpha$ Her) at 9.8, 11.7, \& 18
microns. Our current observations do not confirm any extended Mid-IR
structure around AC Her. These observations are in conflict with
previously reported Keck (seeing-limited) 11.7 and 18 micron images
which suggested the presence of a resolved $\sim 0.6\arcsec$ edge-on
circumbinary disk.  We conclude that AC Her has no extended Mid-IR
structure on scales greater than $0.2\arcsec$ ($R<75$ AU). These first results of Mid-IR AO science are
very encouraging for future high accuracy Mid-IR imaging with this technique.

\end{abstract}

\keywords{instrumentation: adaptive optics --- binaries: general --- stars: evolution --- stars: formation
--- stars: AGB,    Proto-Planetary Nebulae}

\section {Introduction}


Recent evidence suggests that at least a few
post-main sequence giants have acquired long-lived orbiting disks of
dust and molecular gas \citep{jur99}. It has been suggested that these dust disks around post-main sequence stars may possibly
lead to planet formation
\citep{jur00}. It would be quite revolutionary if the process of planet formation occurs at both the ``pre'' and ``post''
main sequence phases of a star's lifetime. Since post--main sequence
stars are much more luminous (L $\sim 1000-10000 L_{\odot}$) than
pre-main sequence stars and have stronger winds and outflows, there
should be very interesting interactions between the disks and these
outflows. Ultimately one might be able to learn more about the
disk/planet formation process if these asymptotic giant branch (AGB)
disks can be confirmed.  Here we report observations of one of the
most interesting of these post-AGB circumstellar disks: AC Her.

The post-AGB spectroscopic binary star AC
Her may be the prototype of these AGB
disk systems. AC Her is one of the most luminous (L$\sim$1000
L$_{\sun}$) and closest (D$\sim$ 750 pc; \cite{she92}) pulsational
variables transiting from the AGB to the
planetary nebula phase (an RV Tauri star). AC Her is a luminous Mid-IR
source ($F_{\nu}$ = 42 Jy at 12 $\mu$m from IRAS) indicating
significant circumstellar dust. \cite{she92} find AC Her's
optical and IR light-curves can be explained by circumstellar dust around a
non-radial pulsator. The origin and nature of this circumstellar dust
has been the subject of speculation. \cite{jur99} argue that the very
narrow (FWHM $\sim 5$ km/s) CO (2-1) emission line
\citep{buj88} in AC Her is a signature of a gravitationally bound (not
outflowing) long-lived reservoir of orbiting gas and
dust. \cite{van98} also argue that there is strong evidence that such a
reservoir of dust may be long-lived and disk-like. \cite{jur99}
further argue that grains in size from 0.2-20 $\mu m$ are present and
grain growth to planetesimal formation is possible. Encouraged by the
sub-mm CO observations of
\cite{buj88} \cite{jur00} obtained 11.7 and 18.7 $\mu m$ images of AC Her at
the 10m Keck telescope in May and August 1999. The
resulting Keck PSF FWHM $\sim 0.35-0.45\arcsec$ (11.7 and 18.7 $\mu m$) images
were the sharpest ever taken of AC Her.

The Keck images of \cite{jur00}
suggest that AC Her is clearly resolved at 18.7 $\mu m$ into north and
south unresolved ``points'' separated by $\sim
0.6\arcsec$ (see our inset in the upper right of Figure \ref{fig1}). \cite{jur00} modeled this image as an edge-on ring of
dust of radius 300 AU in orbit around the 1.39 AU binary AC Her. They
speculated that a binary such as AC Her could produce a small
circumbinary ring of dust which would expand to a radius of 300 AU
over time. They argued that this dust ring would be primarily composed
of long-lived $\sim 200$ $\mu m$ particles which could collide to
create the IR emitting 1 $\mu m$ grains observed in the mid-IR images.
However, since their model relies on the morphology of the Mid-IR
images, it is important to confirm this morphology at higher
Strehls. Hence, we present here very high Strehl Mid-IR images of AC
Her obtained with adaptive optics.


\section{OBSERVATIONS}

 We have utilized the University of Arizona adaptive secondary AO
 system to obtain high resolution images of AC Her
 and several PSF calibration stars. The 6.5 m MMT telescope has a
 unique adaptive optics system. To reduce the aberrations caused by
 atmospheric turbulence all AO systems have a deformable mirror which
 is updated in shape at $\sim 500$ Hz. Until now all adaptive optics
 systems have located this deformable mirror (DM) at a re-imaged pupil
 (effectively a compressed image of the primary mirror). To reimage
 the pupil onto a DM typically requires 6-8 additional warm optical
 surfaces which significantly increases the thermal background and
 decreases the optical throughput of the system
\citep{llo00}. The MMT utilizes a completely new type of DM, which serves as
 both the secondary mirror of the
telescope and the DM of the AO system. 
 In this design, there are no additional optics required in front of
the science camera, the emissivity is lower and thermal IR AO imaging becomes feasible.

The DM consists of 336 voice coil actuators that drive 336 small
magnets glued to the backsurface of a thin (2.0 mm thick) 642 mm
aspheric ULE glass ``shell'' (for a detailed review of the secondary mirror see \cite{bru03a,bru03b}).
 Complete positional control of the surface of this reflective
shell is achieved by use of a capacitive sensor feedback loop. This positional
feedback loop allows one to position an actuator of the shell to
within 4 nm rms (total surface errors amount to only 40 nm rms over
the whole secondary). The AO system samples at 550 Hz using 108 active
subapertures. For a detailed review of the MMT AO system see
\cite{wil03a,wil03b} and references within.

\subsection{MMT Mid-IR AO Observations}

We observed AC Her on the night of 2003, May 13 (UT). The AO system
corrected the lowest 52 system modes and was updated at 550 Hz guiding
on AC Her itself (V=7.03 mag). The closed-loop -3 dB bandwidth was
estimated at 30 Hz. At 1.65 $\mu$m (H band) this level of correction
leads to Strehls of $\sim0.20$ (Close et al. 2003). Since the size
of a coherent patch of air ($r_{\circ}$) increases with
$\lambda^{6/5}$, imaging without AO correction can obtain images
close to diffraction-limited in FWHM once $\lambda > 8 \mu$m. However
such ``seeing-limited'' non-AO Mid-IR images only approach Strehls of
$\sim$0.5 which can lead to significant instability in the PSF
calibration (since approximately half the light is outside the telescope's
diffraction pattern PSF). To increase the Mid-IR Strehl we used AO
correction which vastly improved our AC Her images to nearly perfect
Strehls ($\sim$0.98$\pm0.02$).


\subsubsection{ The MIRAC3 Mid-IR Camera}

We utilized the 128x128 SiAs BIB 2-20$\mu m$ MIRAC3 camera
\citep{hof98}. The $0.088\arcsec$/pixel scale was used with the
9.8, 11.7 and 18 $\mu$m 10\% bandwidth 
filters. To remove thermal and detector instabilities we chopped at
1 Hz with an internal cold chopper in the interface dewar BLINC \citep{hin00} between the AO system and MIRAC3.

We observed with the AO system locked continuously on AC Her.  The
$15^{\circ}$ tilted BLINC dewar window is a high quality dichroic
which reflected the visible light ($\lambda <1$ $\mu$m) to the AO
wavefront sensor and transmitted the IR through BLINC to MIRAC3. Since
the internal chopper in BLINC was past the dichroic, continuous 1 Hz
chopping did not affect the visible light beam and hence the AO lock
was unaffected. To further calibrate the background (in addition
to chopping) we nodded $\sim6-8\arcsec$ in the telescope's azimuth
direction (the horizontal direction in Figure \ref{fig1}) every
minute. The internal chopper was set to run in the altitude direction
(the vertical direction) with a chop throw of $\sim20\arcsec$. The
derotator was disabled during these observations to help minimize the
residual background structure as well.

The $0.505\arcsec$ PA=$269^{\circ}$ Washington Double Star catalog
astrometric binary WDS 02589+2137 BU was observed earlier (2003,
November 25 UT) and used to calibrate the camera's orientation and its
$0.088\arcsec$/pixel platescale.

\subsubsection{Reducing the Mid-IR AO images}

\begin{figure}
 \includegraphics[angle=0,width=\columnwidth]{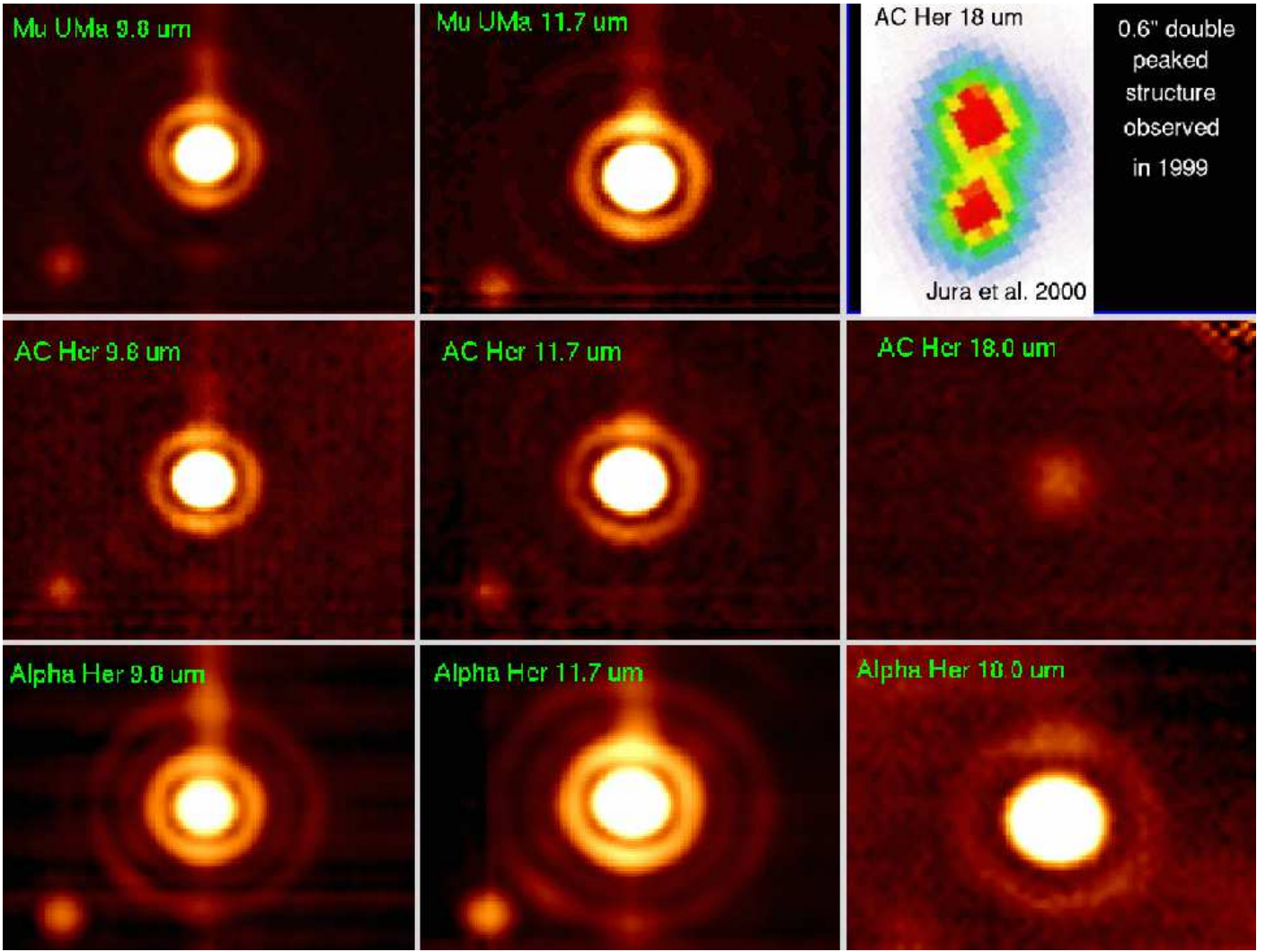}
\caption{
The 9.8, 11.7, and 18 $\mu$m images of AC Her and PSF stars $\mu$ UMa
and $\alpha$ Her as observed at the MMT. In the upper right we have
inserted the published 18 $\mu$m Keck image of AC Her (in false color;
\cite{jur00}). The box size of the MMT images is $1.5x1.0\arcsec$, the effective scale of the Keck image is similar with a box size of $\sim$0.7x1.0$\arcsec$. Note how there is no sign of any extended structure in the MMT AC Her images in any of the filters. The faint point source in the lower left of each MMT image is a MIRAC3 ghost.}
\label{fig1}
\end{figure}

For the 9.8, 11.7, and 18 $\mu$m filters we obtained 8x1 minute coadded chop
differenced images (one image from each nod). Four of these were beam A nods interlaced with 4 beam B nods. We
utilized a custom IRAF script to reduce this Mid-IR data
\citep{bil03}. The script produced eight background subtracted images
by subtracting nod B from the following nod A (and the A nods from the
B nods). The resulting 8 images were bad pixel corrected and
flat-fielded.  The pipeline cross-correlates and aligns each
individual nod image (to an accuracy of $\sim0.02$ pixels), then
rotates each image (by $270^{\circ}$ minus the current parallactic
angle) so north is up and east is to the left. However, there was 
$\le10^{\circ}$ net parallactic rotation for any one filter
observation (over a period of $\sim8$ min.) hence non-rotated images
were also processed on a parallel track. These final aligned images
were median combined. Figure \ref{fig1} illustrates the final AC Her
 images (non-rotated version).

The Mid-IR images of the PSF calibration stars ($\mu$ UMa and $\alpha$
Her; observed before and after AC Her, respectively) were obtained and
reduced in an identical manner to AC Her. In Figure \ref{fig1} we
illustrate our reduced PSF and AC Her images.

\subsection{The PSF Star $\mu$ UMa}

$\mu$ UMa is a well known spectroscopic binary (SB) with a period of
230.089 days and a small eccentricity ($\epsilon$ =0.06;
\cite{bat89}). At an Hipparcos distance of 76.3 pc this suggests an
average separation of $\sim 1.2$ AU. Hence this binary would only
subtend a maximum angle of $0.02\arcsec$ on the sky. Since this is a
factor of 10 less than our resolution limit, the spectroscopic binary
$\mu$ UMa should appear insignificantly different from a point source
with a 6.3m telescope in the Mid-IR. Hence $\mu$ UMa is perfectly
reasonable PSF star for this paper. Moreover, are no reports of
extended Mid-IR structures resolved around $\mu$ UMa to date.

\subsection{The PSF Star $\alpha$ Her}
 
We also utilized $\alpha$ Her as a PSF star. This star is part of a
wide binary system with a fainter (SB) companion located $\sim
4.84\arcsec$ (567 AU) away \citep{fab02}. This companion is not Mid-IR
luminous and was not in our FOV, hence it did not affect our PSF image
of the $\alpha$ Her primary. However, in 1993 the $\alpha$ Her primary
was observed by the ISI interferometer
\citep{dan94} to have a $0.25-0.35\arcsec$ thin shell in the
Mid-IR. We do not detect any evidence of such a shell around the
$\alpha$ Her primary in our Mid-IR AO observations. This is not
surprising given that more recent 1999-2001 ISI measurements also fail
to detect any shell around $\alpha$ Her (S. Tevosian private
communication). Hence, as noted by \cite{wei03}, this shell may have
evolved since the 1993 ISI measurements of \citep{dan94}. It is not
clear how a $R \sim0.25-0.35\arcsec$ (29-41 AU) shell could become
undetectable to the ISI in a period of $\sim 6$ years. Detailed
discussion of the current lack extended structure in the recent ISI
interferometric measurements is beyond the scope of this paper. For
now we simply note that $\alpha$ Her appears to be currently
unresolved (on scales $>0.2\arcsec$) and hence we will utilize it as a
PSF star in this paper.

\section{REDUCTIONS}

\begin{figure}
 \includegraphics[angle=0,width=\columnwidth]{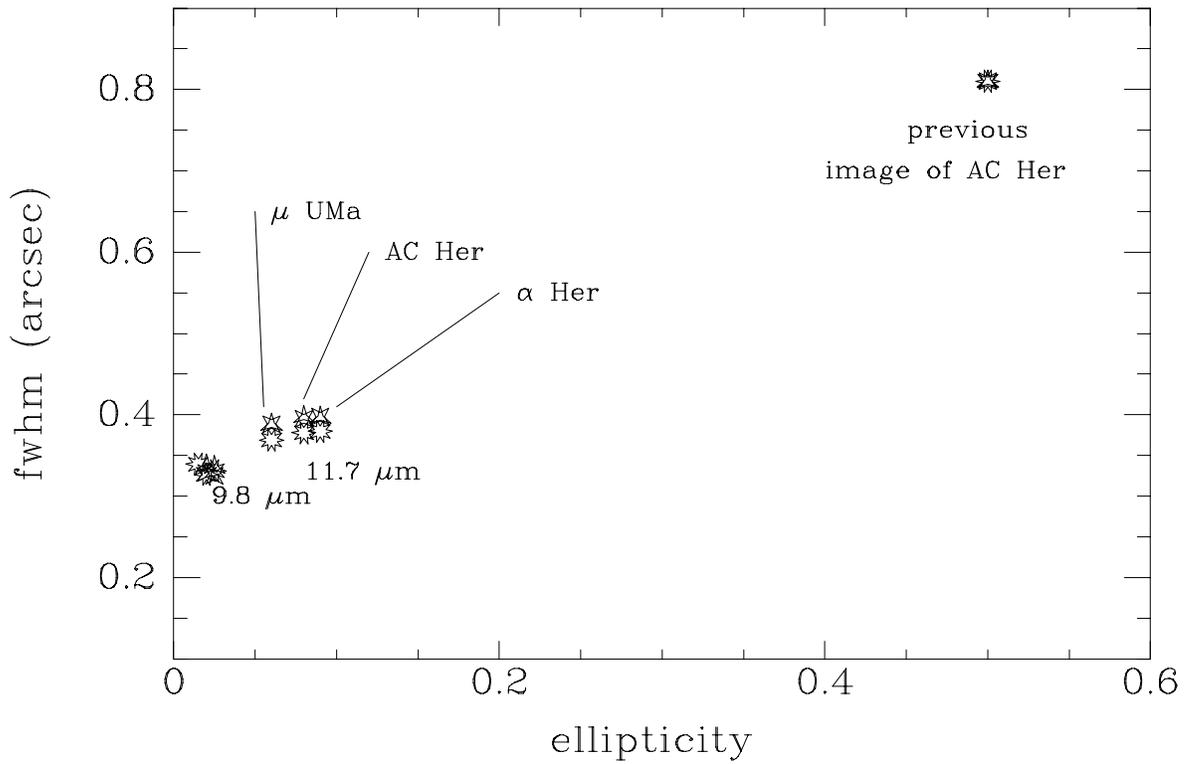}
\caption{
The 9.8 and 11.7 $\mu$m FWHM and ellipticity of AC Her and the PSF
stars $\mu$ UMa and $\alpha$ Her (the Gaussian fit FWHM are the upper
star symbols and the enclosed FWHM are represented by the slightly
lower circles; AC Her is the middle dataset in the 9.8 \& 11.7 $\mu$m
clusters). The location of the ``double-peaked''
 morphology is estimated from the previous Keck image (FWHM $\sim0.8\arcsec$;
\cite{jur00}) to the upper right. Note that AC Her's morphology appears
much more consistent with that of the PSF stars at 9.8 and 11.7 $\mu$m
than an extended FWHM $\sim0.8\arcsec$ disk.}
\label{fig2}
\end{figure}

\begin{figure}
\includegraphics[angle=0,width=\columnwidth]{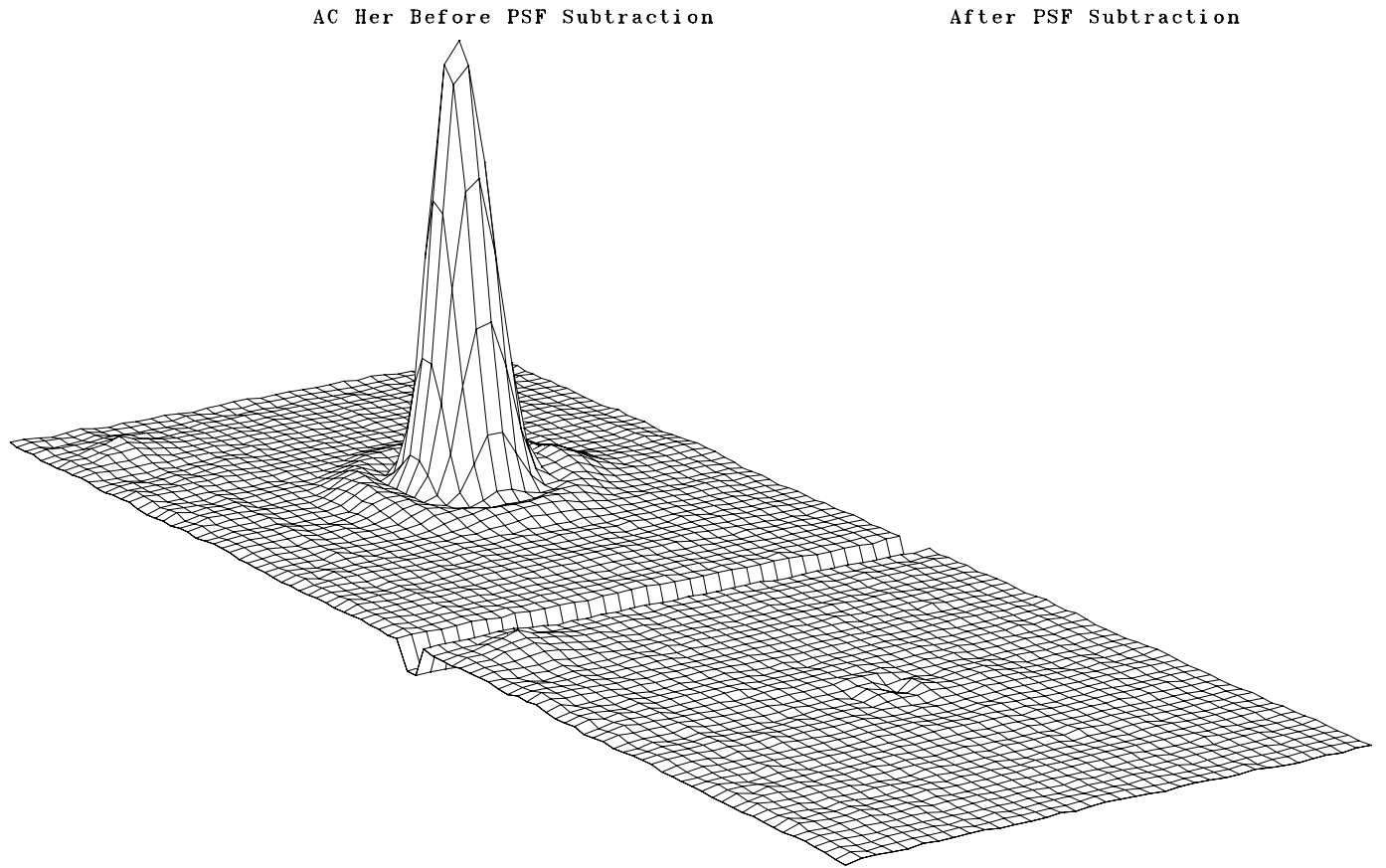}
\caption{
The 11.7 $\mu$m PSF of AC Her before (left) and after (right) PSF
subtraction (using $\alpha$ Her as the PSF) with DAOPHOT's ALLSTAR
task. The residual flux after PSF subtraction is $<0.5\%$ of AC Her's
original flux. Similar residuals resulted from PSF subtractions at 9.8
$\mu$m and 18 $\mu$m. Based on these excellent subtractions it appears
AC Her is not detectably extended. Note that the small ghost image to
the lower left in each frame is not subtracted to show that the
vertical scales are the same for both images.}
\label{fig3}
\end{figure}

We found that the AC Her data appeared consistent with an unresolved
point source. We measured the FWHM, ellipticity, and positional angle
of any such ellipticity for all the images in Figure \ref{fig1}. In
Figure \ref{fig2} we plot the ellipticity and FWHM for our dataset. As
is clear from Figure
\ref{fig2} our AC Her data is very consistent with the PSF stars. Moreover, DAOPHOT's PSF fitting routine ALLSTAR
\citep{ste87} found AC Her to be highly consistent (to within 0.5\%;
see Figure \ref{fig3}) with an unresolved point source ($\alpha$
Her). Hence, it appears that AC Her is point-like in our data.

We also deconvolved AC Her by both the PSF stars $\mu$ UMa and $\alpha$
Her. Due to the very high Strehl ($\sim$0.98) and high
signal-to-noise ratio in our PSF images we could detect low contrast
structure on scales of $\sim 0.2\arcsec$ with the IRAF Lucy
deconvolution task
\citep{bil03}. Even at these small spatial scales we detected no
significant extended structure in the deconvolved images.

\section{ANALYSIS}

As is clear from Figures \ref{fig1} -- \ref{fig3} AC Her is a point
source and is incompatible with the resolved ($0.6\arcsec$ double
peaked) morphology previously observed at Keck and reported by \cite{jur00}. We have confirmed
that we indeed observed AC Her, since the telescope coordinates were
checked twice by an offset from a nearby SAO star. 
The measured 11.7 $\mu$m flux of our object ($\sim$35 Jy) was in
agreement with that of AC Her measured by IRAS (42 Jy at 12
$\mu$m). The
possibility of locking the AO system on another V=7 mag object with $\sim35$
Jy at 11.7 $\mu$m at the location of AC Her (18:30:16.2 +21:52:01
J2000) -- where there are no other nearby 10 $\mu$m sources -- is highly unlikely.

Hence, concluding that we did indeed observe AC Her, it appears
impossible to explain how a long-lived, R$\sim$300 AU circumbinary
disk, could have disappeared since the 1999 observations of
\cite{jur00}. Our deep images (3$\sigma$ $\sim$0.1 Jy) at 9.8 \& 11.7
$\mu$m would have easily detected the $0.6\arcsec$ ``double-peaked''
structure reported by
\cite{jur00}. The BLINC dichroic and pupil lens unfortunately stops transmitting longwards of
$\sim$18$\mu$m, consequently our 18 $\mu$m images have
low throughput and are weighed towards the blue end of the 10\% bandpass filter. However, we should have detected the equal magnitude 
spots even in our
low S/N 18 $\mu$m image, but there is no sign of any $0.6\arcsec$
double-peaked structure at 18 $\mu$m either. 


AC Her's lack of any extended structure subtending a FWHM angle
($\theta_{disk}$) greater than $0\farcs 2$ allows one to place lower
limits on the temperature of dust providing the 41 Jy of flux observed
by IRAS at 12 $\mu$m.  The low line-of-sight optical extinction
(E(B-V)=0.17 mag) to the star and the narrow CO linewidth observations
of \cite{buj88} strongly suggest that the IR flux is produced by a
nearly face-on disk.  Our images imply an upper limit to the disk
diameter of $0\farcs 2$ ($D_{disk} \la 150$ AU), which for optically
thick emission implies a minimum average brightness temperature over
the disk of $\geq200$ K to produce the 12 $\mu$m IRAS flux of 41 Jy.
Simple blackbody emission is clearly inconsistent with the observed
spectral energy distribution (SED).  Our result for the apparent upper
limit to the diameter of the disk at 12 $\mu$m poses a severe
challenge to a satisfactory dust disk model.  In particular, this size
limit is incompatible with the dust model of Shenton et al (1992), who
proposed dust shells of 0$\farcs$5 and 0$\farcs$9 diameter.  Moreover,
the ISO observations of sharp features in the mid-IR spectrum of AC
Her as well as the time variability (Shenton et al. 1992) requires a
significant population of small, warm grains, while the far-IR/mm
spectral index is essentially Rayleigh-Jeans, implying emission from
large cold grains (Jura et al. 2000).  A possible model consistent
with our upper limit on the size is a flared disk of large particles
with a surrounding ``halo" of small particles, as suggested by Jura et
al.  (2000), but either truncated or of such low surface brightness at
12 $\mu$m as to be undetected in our images beyond a diameter of
$0\farcs 2$.  A flared disk model, such as proposed by Jura (2003) for
HD 233517, might offer an explanation for the SED as well as the time
variable IR flux and spectral features.  The interior of the disk will
be a long-lived reservoir of large grains, which is surrounded by a
halo of small grains.  A detailed model calculation for the radiative
transfer is required but beyond the scope of this paper.

\section{CONCLUSIONS}

We find no morphological evidence of any resolved structure around AC
Her. The combination of adaptive optics with a deformable secondary
allows very high Strehl images and high PSF stability regardless of
the seeing, airmass, or target brightness. We are confident that AC
Her appears unresolved in the Mid-IR on scales of $\ge
0.2\arcsec$. This conclusion may impact current theories about whether
or how AGB binaries can produce large (R$\sim$300 AU) long-lived
circumbinary disks since AC Her was the prototypical object. The
hypothesis of a large R$\sim$300 AU circumbinary ring around AC Her
seems unlikely in light of these observations, however, a smaller
$R_{disk}<75$ AU (D$\sim$750 pc) circumbinary disk cannot be ruled out.

Adaptive optics at Mid-IR wavelengths appears to be a very promising
new technique that allows for uniquely stable PSFs and high Strehls. A
high degree of PSF stability will eliminate morphological ambiguities
due to poor (seeing-limited) PSF calibrations. Mid-IR AO should have a
significant impact on any field where Mid-IR imaging is possible.

\acknowledgements

These MMT observations were made possible with the hard
work of the entire Center for Astronomical Adaptive Optics (CAAO)
staff at the University of Arizona. In particular, we would like to
thank Tom McMahon, Kim Chapman, Doris Tucker, and Sherry Weber for
their endless support of this project. The wide field AO CCD was
installed by graduate student Nick Siegler. Dylan Curly helped develop
the MMT AO system user interface. Graduate student Wilson Liu helped
run the MIRAC3 camera during the run. The adaptive secondary mirror is
a joint project of University of Arizona and the Italian National
Institute of Astrophysics - Arcetri Observatory. We would also like
thank the whole MMT staff for their excellent support and flexibility
during our commissioning run at the telescope.

The secondary mirror development could not have been possible without
the support of the Air Force Office of Scientific Research under grant
AFOSR F49620-00-1-0294. LMC acknowledges support from NASA Origins
grant NAG5-12086 and NSF SAA grant AST0206351.  JHB acknowledges
support from NSF grants AST-9987408 and AST-0307687.  We thank the anonymous referee for helpful comments about this paper and our PSF stars. We thank Mike
Jura for helpful discussions and insightful comments.





\end{document}